\documentclass[%
preprint,
amsmath,amssymb,
aps,
floatfix,
superscriptaddress,
showkeys,
]{revtex4-2}
\usepackage{graphicx}
\usepackage{dcolumn}
\usepackage{bm}
\usepackage{hyperref}

\usepackage{multirow}
\usepackage{graphicx}
\usepackage{xcolor}
\usepackage{cancel}

\makeatletter
\renewcommand{\p@subsection}{}
\renewcommand{\p@subsubsection}{}
\makeatother

\makeatletter
\def\@hangfrom@section#1#2#3{\@hangfrom{#1#2}#3}
\def\@hangfroms@section#1#2{#1#2}
\makeatother

\newcommand{\omegatil}{\tilde{\omega}}
\newcommand{\T}{\text}
\newcommand{\mce}{\mathcal{E}}
\newcommand{\mcv}{\mathcal{V}}
\newcommand{\mc}[1]{\mathcal{#1}}
\renewcommand{\vec}[1]{{\boldsymbol{\mathrm{#1}}}}

\begin{document}

\title{Fully Quantum Perturbative Description of Correlated Stokes--anti-Stokes Scattering}

\keywords{Raman scattering, non-linear optics, quantum optics}

\author{Raul Corr\^ea}\email{raulcs@fisica.ufmg.br}
\affiliation{IDOR/Pioneer Science Initiative, Rio de Janeiro, RJ 22281-010, Brazil}
\affiliation{Departamento de Física, Universidade Federal de Minas Gerais, Belo Horizonte, MG 30123-970, Brazil}
\author{Marcelo F. Santos}
\affiliation{Instituto de Física, UFRJ, Caixa Postal 68528, Rio de Janeiro, RJ 21941-972, Brazil}
\author{Carlos H. Monken}
\affiliation{Departamento de Física, Universidade Federal de Minas Gerais, Belo Horizonte, MG 30123-970, Brazil}
\author{Ado Jorio}
\affiliation{Departamento de Física, Universidade Federal de Minas Gerais, Belo Horizonte, MG 30123-970, Brazil}

\date{\today}

\begin{abstract}
The process in which Raman scattering produces correlated Stokes and anti-Stokes radiation is known as Stokes--anti-Stokes (SaS) scattering.
It has been shown recently that this process can generate entangled photon pairs, making it a promising tool for quantum optical technologies, but a proper quantum theoretical description was lacking.
In this paper, a fully quantum derivation of the electric polarization in a medium with vibrational Raman response, with quantized electromagnetic fields, is developed.
Using quantum perturbation theory for Heisenberg operators, we find the solution for the material electric polarization and show that
a four-wave mixing-like correlated SaS scattering appears in the first order of perturbation and completely characterizes the non-resonant SaS photon pair production.
We also discuss how to construct the third-order non-linear optical susceptibility for the SaS scattering from the quantum formalism, and show that it coincides with the one derived for classical fields in stimulated Raman.
\end{abstract}

\maketitle

\section{Introduction}\label{sec:intro}

Raman scattering is a widely explored century-old subject, in which light with frequency $\omega_\ell$ is scattered inelastically by a material degree of freedom mode with frequency $\omegatil$, with the radiation being called Stokes if it is scattered to a lower frequency mode ($\omega_S = \omega_\ell - \omegatil$), and anti-Stokes to a higher one ($\omega_A = \omega_\ell + \omegatil$).
A lesser studied phenomenon is the correlated Stokes--anti-Stokes (SaS) scattering, in which the two scattered modes are coupled.
When an $\omega_S$ scattered mode is propagating in the material along with the incident $\omega_\ell$ mode, their beat frequency is at the material mode, $|\omega_S-\omega_\ell| = \omegatil$, such that the material is stimulated to be excited by the incident light again, giving rise to a coupled $\omega_A$ mode (correlated SaS), or an amplification in the $\omega_S$ mode (stimulated Raman), and vice-versa for a $|\omega_A-\omega_\ell| = \omegatil$ beat.
Historically, this coupling is discussed in the particular context of stimulated Raman with strong fields \cite{bloembergen, foerster71, shen, boyd}, in which an $\omega_S$ strong classical mode is shun on the material along with the $\omega_\ell$ one, in order to create a strong excitation at the beating frequency $\omega_\ell-\omega_S$ and exchange energy with the $\omega_A$ mode.
In \cite{bloembergen}, quantized material degrees of freedom are used to construct the classical third-order non-linear optical susceptibility, while in \cite{shen} a general quantum optical field state is used to calculate the simple Raman transition probability, but the fields are then assumed classical when the SaS coupling is investigated.
In \cite{foerster71}, on the other hand, the quantum dynamics for the Stokes mode optical amplification is solved without explicitly mentioning its interaction with the anti-Stokes mode.

A quantum consequence of the SaS modes coupling is the generation of quantum correlated SaS photon pairs, which was predicted theoretically in 1977 \cite{klyshko1977} and detected experimentally in the last decade \cite{kasperczyk2015, saraiva2017, junior2019, junior2020}, demonstrating even polarization entanglement in the system \cite{freitas2023, freitas2024, freitas2025, vento2025}
and predicting the possibility of squeezed light generation \cite{timsina2024}.
In this phenomenon, both the Stokes and anti-Stokes modes start in the vacuum, and the SaS photon pairs are spontaneously generated, so the stimulated Raman theory for classical light cannot be applied directly.
Nonetheless, the aforementioned references use either a phenomenological theory for the third-order susceptibility of the SaS process, due to the lack of a full quantum optical theory of the phenomenon, or ignore the phonon decay, which cannot explain spectral features.

In this paper we develop a fully quantum non-linear optical theory from first principles, using perturbation theory for Heisenberg operators.
By fully quantum we mean that both matter degrees of freedom and scattered electromagnetic modes are quantized, so we are in the regime of quantum optics.
In Sec. \ref{sec:model}, we construct the vibrational Raman interaction Hamiltonian from first principles.
We use the result in Sec. \ref{sec:perturbation} to solve the dynamics for the material electric polarization with perturbation theory, where it becomes apparent that an important contribution to the correlated SaS scattering comes from the first order of perturbation, and corresponds to a four-wave mixing phenomenon.
In Sec. \ref{sec:chi3}, we construct the third-order susceptibility $\chi^{(3)}$ of the vibrational SaS scattering from our fully quantum theory, and show it coincides with that of the classical theory for stimulated Raman.
In particular, we show how to add a phenomenological electronic third-order susceptibility to obtain the correct SaS photon pair spectrum.
The conclusion is presented in Sec. \ref{sec:conclusion}.

\section{Fully quantum model for vibrational Raman}\label{sec:model}

\begin{figure}[ht]
\begin{center}
    \includegraphics[width=.6\textwidth]{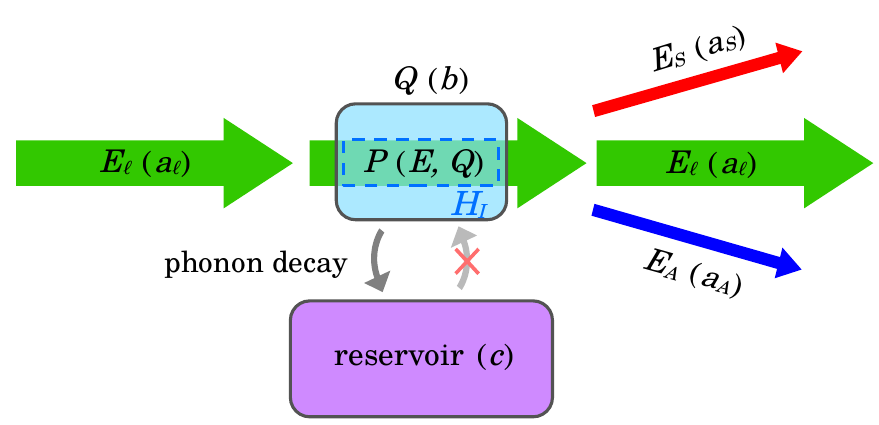}
\end{center}
    \caption{\label{fig:model} Scheme of our theoretical model of vibrational Raman scattering.
        A strong electromagnetic field mode $E_\ell$ (bosonic operators $a_\ell$) interacts with a material whose vibrational degrees of freedom (phonons) are $Q$ (bosonic operators $b$).
        The resulting electric polarization $P(E,Q)$ determines the light-matter interaction, expressed by the interaction Hamiltonian $H_I$.
        The interaction results in the occupation of modes $E_S$ (bosonic operators $a_S$) and $E_A$ (bosonic operators $a_A$), generating the SaS photon pairs.
        At the same time, the Raman active material degrees of freedom $Q$ are coupled to Raman inactive material modes (bosonic operators $c$), which act as a thermal reservoir into which the phonons decay.
    }
\end{figure}

Our model comprises a material whose vibrational degrees of freedom interact with light.
An intense laser field interacts with the material, which gives rise to an electric polarization that produces the scattered light in the Stokes and anti-Stokes modes.
The material vibrations, in turn, are coupled to other material degrees of freedom with which light does not interact, so it causes the phonons to decay.
The overall scheme of our model is in Figure \ref{fig:model}, which we proceed to explain in detail.

The electromagnetic degrees of freedom contain both the laser and the scattered Stokes and anti-Stokes modes, so that our system is fully quantized and closed.
We write the full system Hamiltonian as
\begin{equation}
    H = H_F(t) + H_M(t) + H_I(t),
\end{equation}
where $H_F(t)$ and $H_M(t)$ are the free electromagnetic and material Hamiltonians ($H_M$ includes the phonon decay channel), respectively,
and $H_I(t)$ represents their interaction.
Note that each of them is time-dependent, but since the system is closed (the laser modes are included in $H_F$) their sum $H$ is time-independent.
This is useful, because we can then use the time-independent version $H = H_F(t_0) + H_M(t_0) + H_I(t_0)$, evaluated at the initial time $t_0$, to solve the dynamics of quantum operators \cite{louisell}.

In order to get a vibrational light-matter interaction, we must specify the relation between the material electric polarizability tensor components $\alpha_{i j }(\vec{r},t)$ and the vibrational degrees of freedom, in particular the nuclei displacements $Q_k (\vec{r},t)$ with respect to their equilibrium positions.
The indices $i$, $j$ and $k$ represent spatial directions.
Since vibrational Raman is a very weak interaction, we expand $\alpha_{i j }$ to first order in $Q_k$ \cite{bloembergen, foerster71, boyd},
\begin{equation}\label{eq:polarizability}
    \alpha_{ij}(\mathbf{r},t) = \alpha^{c}_{ij} + \alpha_{ijk} Q_k(\mathbf{r},t),
\end{equation}
where $\alpha^{c}_{ij}$ is an average constant polarizability,
while $\alpha_{ijk} \equiv (\partial \alpha_{ij}/\partial Q_k)|_{Q_k=0}$ is a frequency-independent term related to the strength of polarization fluctuations due to material vibrations, and it is related to the Raman polarizability tensor \cite{levenson74}.

The polarizability $\alpha_{ij}$ characterizes the electric polarization component $P_i$ in the material as an instantaneous response to an electric field $E_j$,
\begin{equation}\label{eq:pol_response}
    P_i (\vec{r}, t)
    =
    N \epsilon_0 \alpha_{ij}(\vec{r},t)
    E_j (\vec{r}, t),
\end{equation}
where $\epsilon_0$ is the vacuum electric permittivity,
$N$ is the number of dipoles per unit volume responding to the electric field,
and repeated latin indices summation is implied here and throughout the article.

When we insert the polarizability expansion (\ref{eq:polarizability}) in the polarization expression (\ref{eq:pol_response}), we get a term proportional only to $E_j (\vec{r},t)$, and another term proportional to the product $E_j (\vec{r},t) Q_k(\vec{r},t)$.
The former is just the effect of a refractive index, while the latter is the Raman electric polarization.
The refractive index part can be incorporated into the electromagnetic Hamiltonian $H_F(t)$ \cite{foerster71, raymer2020}, so only the Raman part of the polarization,
\begin{equation}\label{eq:pol_raman}
    P^\T{R}_i  (\vec{r},t)
    \equiv
    N\epsilon_0 \alpha_{ijk}
    E_j  (\vec{r},t)
    Q_k (\vec{r},t),
\end{equation}
is included in the light-matter interaction Hamiltonian $H_I(t)$, leading to
\begin{equation}\label{eq:H_I_proto}
    H_I(t)
    =
    \frac{N\epsilon_0 \alpha_{ijk}}{2}
    \int_\mcv E_i (\vec{r},t) E_j (\vec{r},t) Q_k(\vec{r},t) d^3{\vec{r}} ,
\end{equation}
where $\mcv$ is the domain of the interaction volume.

The Hamiltonian in (\ref{eq:H_I_proto}) can be further simplified under the assumption that the laser spectrum does not overlap with that of the scattered modes,
which is true whenever the differences between the scattered and laser frequencies are larger than the laser bandwidth.
With that assumption, as well as considering a very intense laser when compared with the weak scattered fields \cite{foerster71}, we have
\begin{equation}\label{eq:H_I}
    H_{I}(t)
    =
    N\epsilon_0 \alpha_{ijk}
    \int_{\mc{V}} E_i (\vec{r},t) \mce_j  (\vec{r},t) Q_k (\vec{r},t) d^3{\vec{r}},
\end{equation}
where we used $\mce_j $ to represent the laser modes and reserve $E_i$ for the scattered modes.

The Hamiltonian for the electromagnetic modes is
\begin{equation}
    H_F(t) = \sum_{\vec{k}, \sigma}
    \hbar \omega(\vec{k}) [
        a_{\vec{k} \sigma}^\dagger (t) a_{\vec{k} \sigma} (t)
        + 1/2
    ],
\end{equation}
where $a_{\vec{k} \sigma}$ are annihilation operators of a light mode with wave vector $\vec{k}$ and polarization $\sigma$,
describing either a laser or a scattered mode.
The mode frequencies $\omega(\vec{k})$ are defined by the relation
$|\vec{k}| = \omega n(\omega)  / c$, where $n(\omega)$ is the medium refractive index \cite{raymer2020}.
They are quantum bosonic operators with commutation relations
$[ a_{\vec{k} \sigma} (t), a_{\vec{k}' \sigma'}^\dagger (t) ] = \delta_{\vec{k}, \vec{k}'} \delta_{\sigma, \sigma'}$ and $[ a_{\vec{k} \sigma} (t), a_{\vec{k}' \sigma'} (t) ] = 0 $.
The electric fields relate to the plane wave bosonic operators via
\begin{equation}\label{eq:E_and_a}
    E_i  (\vec{r}, t)
    =
    \sum_{\vec{k}_s, \sigma}
        \sqrt{\frac{\hbar \omega_s}{2 \epsilon_0 n^2(\omega_s) V_Q}}
        \Big[
        i a_{\vec{k}_s\sigma}(t) e^{i\vec{k}_s\cdot\vec{r}} \varepsilon_{\vec{k}_s \sigma i }
        + \T{H.c.}
        \Big],
\end{equation}
where $V_Q$ is the quantization volume, $\varepsilon_{\vec{k}_s \sigma i }$ is the projection of the unit polarization vector of mode $(\vec{k}_s,\sigma)$ on direction $i $,
and $\T{H.c.}$ denotes the Hermitian conjugate of the previous term.
We use $\vec{k}_s$ to denote the scattered modes, so an analogous expression exists for the laser field $\mce_j (\vec{r},t)$, in which we will use $\vec{k}_\ell$ to represent its modes.
When there is no Raman interaction ($H_I(t) = 0$), the evolution of the photon operators for either laser or scattered modes is straight-forward,
$\bar{a}_{\vec{k} \sigma}(t) = a_{\vec{k} i}(t_0) e^{-i \omega (t-t_0)}$,
where we use a bar to denote the evolution with the interaction turned off.

Until now we have been concerned with the light-matter interaction, but we still need to specify how the phonons decay.
To do that with a quantum model, we use a phonon reservoir coupled to the Raman active phonons, such that the reservoir does not interact with light directly.
We thus write the material Hamiltonian as
\begin{eqnarray}\label{eq:H_M}
    H_M (t)
    &=&
    \sum_{\vec{q}, \eta} \hbar \omega_{0\eta}[ b_{\vec{q} \eta}^{\dagger} (t) b_{\vec{q} \eta} (t) + 1/2]
    + \sum_{\vec{q}, r, \eta} \hbar \omega_{r\eta}[ c_{\vec{q}r \eta}^{\dagger} (t) c_{\vec{q}r \eta} (t) + 1/2] \\
    && + \sum_{\vec{q}, r, \eta} \hbar [\zeta_{r\eta}^* c_{\vec{q}r\eta}^{\dagger} (t) b_{\vec{q} \eta} (t) + \zeta_{r\eta} c_{\vec{q}r\eta} (t) b_{\vec{q} \eta}^{\dagger} (t) ], \nonumber
\end{eqnarray}
where $b_{\vec{q}\eta}$ are annihilation operators of a Raman active phonon mode with wave vector $\vec{q}$ and polarization $\eta$,
and $c_{\vec{q}r\eta}$ is the counterpart for a phonon reservoir mode with wave vector $\vec{q}$, energy index $r$, and polarization $\eta$.
The operators $b_{\eta \vec{q}} (t)$ and $c_{\eta r \vec{q}} (t)$ are quantum bosonic operators
with commutation relations $[ b_{\vec{q} \eta} (t), b_{\vec{q}' \eta'}^\dagger (t) ] = \delta_{\vec{q}, \vec{q}'} \delta_{\eta, \eta'}$, $[ b_{\vec{q} \eta} (t), b_{\vec{q}' \eta'} (t) ] = 0 $,
$[ c_{\vec{q} \eta r} (t), c_{\vec{q}' \eta' r'}^\dagger (t) ] = \delta_{\vec{q}, \vec{q}'} \delta_{r, r'} \delta_{\eta, \eta'}$ and $[ c_{\vec{q} \eta r} (t), c_{\vec{q}' \eta' r'} (t) ] = 0 $.
The frequencies $\omega_{0\eta}$ and $\omega_{r\eta}$ are the free oscillator frequencies of the phonon and reservoir modes, respectively,
and $\zeta_{r\eta}$ quantifies their interaction.
Note that we assume a flat dispersion relation for the phonons, which is reasonable for optical phonons near $\vec{q} = 0$ \cite{foerster71}.
The relation between $b_{\vec{q}\eta}(t)$ and $Q_k(\vec{r},t)$ is
\begin{equation}\label{eq:Q_and_b}
    Q_{k} (\vec{r}, t)
    =
    \sum_{\vec{q}, \eta}
        \sqrt{\frac{\hbar}{2 M_\eta \omega_{0\eta}}}
        \Big[
        b_{\vec{q}\eta}(t) e^{i\vec{q}\cdot\vec{r}} \varepsilon_{\vec{q} \eta k}
        + \T{H.c.}
        \Big],
\end{equation}
where $M_\eta$ is the effective mass of the oscillator and $\varepsilon_{\vec{q} \eta k}$ is the projection of the unit polarization vector of mode $(\vec{q},\eta)$ on direction $k$.

The solution of the dynamics for $\bar{b}_{\vec{q}\eta}(t)$, subject only to the material Hamiltonian, i.e. without the Raman interaction ($H_I(t) = 0$), can be obtained by assuming that there are infinite reservoir modes and the Weisskopff-Wigner approximation \cite{louisell,vankampen,guimaraes2020}.
In this approximation, the excitations decay to the reservoir and do not go back to the Raman mode.
The solution contains a transient term, oscillating with the complex frequency $\omegatil_{\eta} -i \gamma_\eta/2$
(whose real part is the Raman phonon frequency shifted by the coupling, and the imaginary part is its decay rate)
and a stationary term, oscillating with the real frequency $\omega_{r\eta}$.
In the solution expression below, the transient term is the one accompanying the square brackets, whereas $C_{\vec{q} \eta} (t)$ is the stationary term,
\begin{equation}\label{eq:phonon_free_solution}
    \bar{b}_{\vec{q} \eta}(t)
    =
    \Big[
        b_{\vec{q} \eta}(t_0)
        -
        C_{\vec{q} \eta} (t_0)
    \Big]
        e^{-i( \omegatil_{\eta} -i \gamma_\eta/2 ) (t-t_0)}
    + C_{\vec{q} \eta} (t),
\end{equation}
where
\begin{equation}
    C_{\vec{q} \eta} (t)
    \equiv
    \sum_r \zeta_{r\eta} \frac{c_{\vec{q}r\eta} (t_0) e^{-i\omega_{r\eta} (t-t_0)} }{\omega_{r\eta} -\omegatil_\eta +i\gamma_\eta/2}
    .
\end{equation}
The transient part accounts for the $b-c$ dynamics, in which the energy is still being transferred from the Raman phonon to the reservoir,
while the stationary part ($C_{\vec{q} \eta} (t)$) contains the dynamics of the phonon already lost to the reservoir.

As a last step in our preparation of the fully quantum model of the Raman scattering, we develop the interaction Hamiltonian (\ref{eq:H_I}) with plane wave mode operators for the electric field and for the material vibrations,
\begin{subequations}\label{eq:plane_wave}
    \begin{equation}
    E_{\vec{k}_s i }(t)
    =
    \sum_{\sigma_s} i \sqrt{\frac{\hbar \omega_s}{2 \epsilon_0 n^2(\omega_s) V_Q}}
    a_{\vec{k}_s \sigma_s}(t)
    \varepsilon_{\vec{k}_s \sigma_s i },
    \end{equation}
    \begin{equation}
    Q_{\vec{q} k}(t)
    =
    \sum_\eta \sqrt{\frac{\hbar}{2 M_\eta \omega_{0\eta}}}
    b_{\vec{q} \eta}(t)
    \varepsilon_{\vec{q} \eta k},
    \end{equation}
\end{subequations}
and an analogous expression for the laser modes $\mce_{\vec{k}_\ell j }(t)$.
Using Equations (\ref{eq:E_and_a}), (\ref{eq:Q_and_b}) and (\ref{eq:plane_wave}) in the Hamiltonian of Equation (\ref{eq:H_I}), there will be many products of operators and its conjugates, like
$E_{\vec{k}_s i } \mce_{\vec{k}_\ell j } Q_{\vec{q} k}$,
$E_{\vec{k}_s i } \mce_{\vec{k}_\ell j } Q_{\vec{q} k}^\dagger$,
$E_{\vec{k}_s i } \mce_{\vec{k}_\ell j }^\dagger Q_{\vec{q} k}$,
\dots
The interaction is weak, so the time evolution of these operators will be a slowly varying amplitude times a fast oscillation with frequency associated with the respective plane wave mode,
$\omega_\ell \equiv \omega_\ell(\vec{k}_\ell)$,
$\omega_s \equiv \omega_s(\vec{k}_s)$,
and
$\omega_p \equiv \omega_p(\vec{q})$.
The phonon frequency $\omega_p$ can be either a resonance frequency $\omegatil_\eta$ or a reservoir frequency $\omega_{r\eta}$ close to $\omegatil_\eta$.
Each operator product will thus accompany an exponential factor of the kind
$\exp[-i(\pm \omega_\ell \pm \omega_s \pm \omega_p)t]$,
where we associate positive signs with the annihilation operators and negative with the creation ones.
Only when $(\pm \omega_\ell \pm \omega_s \pm \omega_p) \approx 0$ does the exponential not oscillate in time, so after some cycles only the frequency combinations close to zero will survive,
which is a rotating wave approximation.
Because the phonon frequencies are much smaller than the optical field frequencies used in the experiments, the approximation will yield $\omega_p \approx |\omega_\ell -\omega_s|$,
which can be interpreted as an energy conservation condition, associated with temporal phase matching.
Also, it means that only combinations of one annihilation and one creation operator for photons will survive.

On top of that, for each time exponential there will be a space counterpart
$\exp[i(\pm \vec{k}_\ell \pm \vec{k}_s \pm \vec{q}) \cdot \vec{r}]$
which, after integration over the scattering volume, will result in delta factors provided that the medium dimensions are much larger than the typical wavelengths involved,
since $\int_{\mcv} e^{i(\vec{k}-\vec{k}')\cdot \vec{r}} d^3\vec{r} \approx V_S \delta_{\vec{k}, \vec{k}'}$, where $V_S$ is the total scattering volume over $\mcv$.
Analogously, this is interpreted as a momentum conservation condition, associated with spatial phase matching.
If finite medium dimensions are considered, then momentum conservation is not exact \cite{foerster71}.
Combining plane wave modes with energy and momentum conservation, the Raman interaction Hamiltonian, Equation (\ref{eq:H_I}), is rewritten as
\begin{equation}
    H_{I}(t)
    =
    N\epsilon_0 \alpha_{i  j  k}
    \sum_{\vec{k}_\ell} \sum_{\vec{q}}
    \bigg\{
        \Big[
            \mce_{\vec{k}_\ell j } (t) E^{\dagger}_{(\vec{k}_\ell-\vec{q}) i }(t) Q_{\vec{q} k}^\dagger (t)
            +
            \mce_{\vec{k}_\ell j } (t) E^{\dagger}_{(\vec{k}_\ell+\vec{q}) i }(t) Q_{\vec{q} k} (t)
        \Big]
         + \T{H.c.}
    \bigg\} ,
\end{equation}
where the first (second) term between square brackets is responsible for the creation of photons in the Stokes (anti-Stokes) mode, as is evident by the phonon operators and the resulting wave vector of the created photons.

\section{Perturbative solution for the electric polarization}\label{sec:perturbation}

The electric polarization is the central physical quantity that contains all the field+matter interaction dynamics in response to the incident field, so we need to calculate its evolution.
In particular, we are interested in the Raman part of the polarization, $P^\T{R}_i  (\vec{r},t)$ of Equation (\ref{eq:pol_raman}).
Since our system is closed, the equation we have to solve can be written as
\begin{equation}
    \frac{d P^\T{R}_i }{d t} (\mathbf{r},t) = -\frac{i}{\hbar} [P^\T{R}_i  (\mathbf{r},t), H_F +H_M +H_I ],
\end{equation}
in which the operators without the time dependence indication are calculated at the time in which the interaction starts, $t = t_0$.
Then we can use a perturbative expansion of the solution for the operators dynamics \cite{louisell},
\begin{equation}
    P^\T{R}_i (\mathbf{r},t) = \bar{P}^\T{R}_i (\mathbf{r},t)
    -\frac{i}{\hbar}\int_{t_0}^t [\bar{P}^\T{R}_i (\mathbf{r},t), \bar{H}_I(t-t')] d t' +\dots,
\end{equation}
where the bar over an operator $O(t)$ denotes its interaction representation
$$\bar{O}(t) \equiv e^{i H_0 (t-t_0)/\hbar} O(t) e^{-i H_0 (t-t_0)/\hbar},$$
with $H_0 = (H_F+H_M)$ as the free Hamiltonian at the initial time $t_0$.
All field modes, incident and scattered, must be included in the $\bar{P}_i^{\T{R}}$ and $\bar{H}_I$ operators, but some of the resulting operator combinations will describe the evolution of the incident field modes, i.e. how the laser gains or loses photons.
However, our perturbation theory only adds or subtracts a small number of photons from the modes, and since we consider the incident laser to be a strong field, with a number of photons much larger than one, the modifications in the laser state can be neglected, and we will only write the solution terms that generate Stokes and anti-Stokes photons.

The zeroth order of perturbation will give us the polarization that generates independent Stokes and anti-Stokes modes,
\begin{equation}\label{eq:pol_zeroth}
    P^{\T{R} (0)}_i (\mathbf{r},t) = N\epsilon_0 \alpha_{ijk}
    \bar{\mathcal{E}}_j  (\mathbf{r},t)
    \bar{Q}_k(\mathbf{r},t).
\end{equation}
Writing it in terms of field plane wave modes
we can explicitly identify the Stokes and anti-Stokes production terms,
\begin{equation}\label{eq:pol_zeroth_plane}
    P^{\T{R} (0)}_i (\vec{r},t)
    =
    N\epsilon_0 \alpha_{i  j  k}
     \sum_{\vec{k}_\ell} \sum_{\vec{q}} \left\{
        \left[
        \bar{\mce}_{\vec{k}_\ell j }(t)
        \bar{Q}_{\vec{q} k}^\dagger (t)
        e^{i (\vec{k}_\ell-\vec{q}) \cdot \vec{r}}
        +
        \bar{\mce}_{\vec{k}_\ell j }(t)
        \bar{Q}_{\vec{q} k} (t)
        e^{i (\vec{k}_\ell+\vec{q}) \cdot \vec{r}}
        \right] + \T{H.c.} \right\},
\end{equation}
which after plugging the free solutions for the $\bar{\mce}$ and $\bar{Q}$ operators (see (\ref{eq:phonon_free_solution})) leads to an expression analogous to the polarization source term calculated in \cite{guimaraes2020}.

In the first order of perturbation there is a commutator involving the laser and vibration modes at different times.
At this point, for the sake of simplicity, we can assume that the laser is intense enough to treat its operators as numbers (i.e. $\langle \mce\mce^\dagger \rangle \approx \langle \mce^\dagger\mce \rangle$), so the commutator only involves phonon operators.
It yields
\begin{equation}\label{eq:pol_first}
    P^{\T{R} (1)}_i (\mathbf{r},t) = -\frac{i}{\hbar} (N\epsilon_0)^2 \alpha_{ijk} \alpha_{i'j' k'}
    \int_{t_0}^t d t' \int_\mathcal{V} d^3\mathbf{r}'
    [\bar{Q}_{k}(\mathbf{r},t), \bar{Q}_{k'}(\mathbf{r}',t-t')]
    \bar{\mathcal{E}}_j  (\mathbf{r},t)
    \bar{\mathcal{E}}_{j'}(\mathbf{r}',t-t')
    \bar{E}_{i'}(\mathbf{r}',t-t'),
\end{equation}
which is a four-wave mixing (FWM) interaction, involving three electric fields.
Importantly, note that the commutator of quantum operators is a number, so the first-order term does not depend on the material state.
This means that the FWM scattering contains the correlated SaS instantaneous interaction, in which no real phonons are created or annihilated, and only virtual instantaneous transitions are involved, even on resonance.
Matter vibrations then act as a mediator between the four fields, leading to a third-order susceptibility.

We can calculate $[\bar{Q}_{k}(\mathbf{r},t), \bar{Q}_{k'}(\mathbf{r}',t-t')]$ with the help of $[ \bar{b}_{\vec{q} \eta} (t), \bar{b}_{\vec{q}' \eta'}^\dagger (t-t')]$, which is obtained from the solution $\bar{b}_{\vec{q}\eta}(t)$, Equation (\ref{eq:phonon_free_solution}),
and the commutation relations for $b_{\vec{q} \eta} (t_0)$ and $c_{\vec{q} \eta} (t_0)$.
Details of the calculation can be found in Appendix \ref{app:commutator}, which yields
\begin{equation}\label{eq:b_commutator}
	[ \bar{b}_{\vec{q} \eta} (t), \bar{b}_{\vec{q}' \eta'}^\dagger (t-t')]
    =
    e^{-i \omegatil_\eta t'}
    e^{-\gamma_\eta |t'| /2}
    \delta_{\vec{q},\vec{q}'}
    \delta_{\eta,\eta'}.
\end{equation}

With that result, one can write the first-order polarization operator (\ref{eq:pol_first}) as
\begin{eqnarray}\label{eq:pol_fAS}
    P_i ^{\T{R} (1)}(\mathbf{r},t)
    &=&
	\sum_\eta \frac{(N\epsilon_0)^2}{2 M_{\eta} \omega_{\eta 0}}
    \alpha_{i  j  k} \alpha_{i'j'k'}
	\sum_{\vec{k}_\ell} \sum_{\vec{q}}
	\sum_{\vec{k}_\ell'}
	\bigg\{
    \nonumber
        \\
    &&\mkern-50mu
    \bigg[
        \bar{\mce}_{\vec{k}_\ell j } \bar{\mce}_{\vec{k}_\ell' j'}
	\bar{E}_{(\vec{k}_\ell'-\vec{q}) i'}^\dagger
		e^{i(\vec{k}_\ell+\vec{q})\cdot\vec{r}}
        \varepsilon_{\vec{q} \eta k'}^* \varepsilon_{\vec{q} \eta k}
    f_{++-}^{AS}(t)
    - \bar{\mce}_{\vec{k}_\ell j } \bar{\mce}_{\vec{k}_\ell' j'}
    \bar{E}_{(\vec{k}_\ell'+\vec{q}) i'}^\dagger
		e^{i(\vec{k}_\ell-\vec{q})\cdot\vec{r}}
        \varepsilon_{\vec{q} \eta k'} \varepsilon_{\vec{q} \eta k}^*
    f_{++-}^{SA}(t)
    \nonumber\\
    &&\mkern-40mu
    + \bar{\mce}_{\vec{k}_\ell j } \bar{\mce}_{\vec{k}_\ell' j'}^*
    \bar{E}_{(\vec{k}_\ell'+\vec{q}) i'}
		e^{i(\vec{k}_\ell+\vec{q})\cdot\vec{r}}
        \varepsilon_{\vec{q} \eta k'}^* \varepsilon_{\vec{q} \eta k}
    f_{+-+}^{AA}(t)
    - \bar{\mce}_{\vec{k}_\ell j } \bar{\mce}_{\vec{k}_\ell' j'}^*
    \bar{E}_{(\vec{k}_\ell'-\vec{q}) i'}
		e^{i(\vec{k}_\ell-\vec{q})\cdot\vec{r}}
        \varepsilon_{\vec{q} \eta k'} \varepsilon_{\vec{q} \eta k}^*
    f_{+-+}^{SS}(t)
    \bigg]
    \nonumber\\
    &&\mkern-50mu
    + \T{H.c.}
	\bigg\}.
\end{eqnarray}
We set for convenience $t_0=0$, which yields
\begin{equation}\label{eq:f_time}
    f_{\pm \pm \pm}^{(\frac{S}{A}) (\frac{S}{A})} (t)
    =
    f_{\pm \pm \pm}^{(\frac{S}{A}) (\frac{S}{A})} (t; \vec{k}_\ell, \vec{k}_\ell', \vec{q})
    =
    \frac{
        e^{-i(\pm\omega_\ell \pm\omega_\ell' \pm\omega_{\ell\mp}')t}
        - e^{-i(\pm\omega_\ell \mp\omegatil_\eta -i\gamma_\eta/2)t}
    }
    {( \pm \omega_{\ell}' \pm \omega_{\ell\mp}' \pm \omegatil_\eta + i \gamma_\eta/2
    )},
\end{equation}
where $\omega_\ell \equiv \omega_\ell(\vec{k}_\ell)$,
$\omega_\ell' \equiv \omega_\ell(\vec{k}_\ell')$
and $\omega_{\ell\pm}' \equiv \omega_s(\vec{k}_\ell' \pm \vec{q})$
are the frequencies associated with the plane wave modes of the electric fields.
The three signs in the subscript of $f$ indicate whether the three waves (laser fields $\bar{\mce}_{\vec{k}_\ell}$ and $\bar{\mce}_{\vec{k}_\ell'}$, and scattered field $\bar{E}_{\vec{k}_s}$, respectively) have a positive or a negative frequency in the associated term.
The first $S/A$ superscript indicates which electric polarization mode, whether a Stokes or anti-Stokes, does the term refer to (check it in the spatial exponential,
e.g. $e^{-i(\vec{k}_\ell\mp\vec{q})\cdot\vec{r}}$,
with minus for Stokes and plus for anti-Stokes),
and the sign of the phonon frequency $\pm\omegatil_\eta$ is associated with it
(top sign for Stokes and bottom one for anti-Stokes).
The second $S/A$ superscript indicates to which other scattered mode is the term correlated with (check it in the scattered mode field operator,
e.g. $\bar{E}_{\vec{k}_\ell'\mp\vec{q}}$,
with minus for Stokes and plus for anti-Stokes).
The subscript sign of $\omega_{\ell\mp}'$ then takes the top value for correlated Stokes (second superscript $S$) and the bottom one for correlated anti-Stokes (second superscript $A$).

As already stated, the expression for the first-order solution of the material electric polarization (\ref{eq:pol_fAS}) is in the form of a FWM non-linear optical phenomenon, in which there are always three fields on the right side of the equation which, combined, lead to the polarization that generates the fourth field.
For instance, in the first line, the first term contains the field of two laser modes, $\vec{k}_\ell$ and $\vec{k}_\ell'$, being absorbed and creating one photon in mode $\vec{k}_\ell'-\vec{q}$ (Stokes), while the polarization is in mode $\vec{k}_\ell+\vec{q}$ (anti-Stokes).
This term is thus associated with the scattering from $\vec{k}_\ell'$ into a Stokes mode, and $\vec{k}_\ell$ into an anti-Stokes mode.
The second term in the first line is analogous, but $\vec{k}_\ell$ scatters into a Stokes mode, $\vec{k}_\ell-\vec{q}$, and $\vec{k}_\ell'$ into an anti-Stokes one, $\vec{k}_\ell'+\vec{q}$.
The first line therefore contains the Stokes--anti-Stokes coupling terms, in which these modes are correlated.

In the second line, we have the self-coupling terms, since in the first (second) term the absorbed modes are one from the laser, $\vec{k}_\ell$, and one anti-Stokes (Stokes) mode $\vec{k}_\ell'+\vec{q}$ ($\vec{k}_\ell'-\vec{q}$), coupling with the emission back into the laser mode $\vec{k}_\ell'$ and the anti-Stokes (Stokes) polarization $\vec{k}_\ell+\vec{q}$ ($\vec{k}_\ell'-\vec{q}$).
They are associated with stimulated emission.

Importantly, note that no phonon transition operators $\bar{Q}_{\vec{q} k} (t)$ appear in Equation (\ref{eq:pol_fAS}), so the FWM part of the solution leads to the same result whatever the initial material state is, only causing transitions on electromagnetic degrees of freedom.
In other words, it leaves the material unchanged and it is insensitive to the quantum state of the material.
This tells us that the FWM contribution to the SaS scattering is not affected by the temperature of the material, like simple Raman is (see Equation (\ref{eq:pol_zeroth_plane})),
but the medium only passively mediates the FWM interaction between the electromagnetic field modes.

One must recall, however, that there are zeroth-order contributions to the SaS scattering, since the generation of independent Stokes and anti-Stokes photons in simple Raman can contribute with SaS pairs, though they do not carry the kind of correlation that FWM-generated pairs do (e.g. those studied by \cite{freitas2023, freitas2024, freitas2025, vento2025}).
These zeroth-order contributions, as evident in Equation (\ref{eq:pol_zeroth_plane}), do contain $\bar{Q}_{\vec{q} k} (t)$ operators and are sensitive to the material quantum state, and therefore to its temperature.

\section{Discussion}\label{sec:chi3}

Having calculated the non-linear polarization as a function of the three fields that are associated with it, we can derive the third-order susceptibility for the vibrational Raman scattering $\chi^{(3) \T{R}}_{ijj'i'} (-\omega_A, \omega_\ell, \omega_\ell', -\omega_S)$,
given by
\begin{equation}
    P^{\T{R}(1)}_i(\omega_A)
    =
    \sum_{\omega_\ell} \sum_{\omega_\ell'} \sum_{\omega_S}
    \chi^{(3) \T{R}}_{ijj'i'} (-\omega_A, \omega_\ell, \omega_\ell', -\omega_S)
    \mc{E}_{\vec{k}_\ell j}(\omega_\ell) \mc{E}_{\vec{k}_\ell' j'}(\omega_\ell')
	E_{\vec{k}_S i'}^\dagger (\omega_S)
    .
\end{equation}
However, because our theory is fully spatio-temporal, we have to go through a spatial third-order susceptibility in order to derive the usual temporal frequency susceptibility,
and the calculation can be found in Appendix \ref{app:susceptibility}.
The susceptibility will be proportional to
$f_{\pm \pm \pm}^{(\frac{S}{A}) (\frac{S}{A})} (\omega)
\equiv \int_{-\infty}^{\infty}
f_{\pm \pm \pm}^{(\frac{S}{A}) (\frac{S}{A})} (t)
e^{i\omega t} dt$,
that is, the Fourier transform of the function in Equation (\ref{eq:pol_fAS}) mediating the three fields and the resulting non-linear polarization,
with the first exponential surviving while the second one goes to zero because we are integrating over long times.

We write it as
\begin{equation}\label{eq:chi3_R}
    \chi^{(3) \T{R}}_{ijj'i'} (-\omega_A, \omega_\ell, \omega_\ell', -\omega_S)
    =
    A^{\T{R}}_{ijj'i'}
    \frac{ \gamma }
    {( \omega_{\ell}' - \omega_S - \omegatil + i \gamma/2
    )}
    \delta( \omega_{\ell} + \omega_{\ell}' - \omega_{S} -\omega_A)
    ,
\end{equation}
where
$A^{\T{R}}_{ijj'i'} \equiv
    \sum_\eta
    \bar{\alpha}_{ij,\eta} \bar{\alpha}_{i'j',\eta}^*$,
and
$\bar{\alpha}_{ij,\eta}(\vec{q}) \equiv
    \sqrt{\frac{2\pi}{\gamma_\eta}}
    \frac{V_S}{2 M_{\eta} \omega_{\eta 0}}
    N\epsilon_0
    \alpha_{ijm}
    \varepsilon_{\vec{q} \eta m}$
can be taken as independent of $\vec{q}$ if the scattering is over small angles.
Equation (\ref{eq:chi3_R})
is the susceptibility due to the FWM Raman interaction in the material,
for the $\omega_\ell'$ mode scattering into $\omega_S$, and $\omega_\ell$ into $\omega_A$.
    It has a resonance when the Raman shift equals the phonon frequency, $\omega_\ell'-\omega_S = \omegatil$, appearing in the Lorentzian probability amplitude, and energy conservation between the four photons demands that $\omega_\ell+\omega_\ell' = \omega_S+\omega_A$, which appears in the Dirac delta.
It has the same functional form as the susceptibility calculated for stimulated Raman scattering \cite{bloembergen,boyd}, but since we have made a fully quantum treatment, we have shown that this expression can be used even in the spontaneous SaS photon pair generation, as it has been without a formal derivation \cite{sier2024,vento2025}.

In a perturbative quantum field theory, the third-order susceptibility $\chi^{(3)}_{ijj'i'} (-\omega_A, \omega_\ell, \omega_\ell', -\omega_S)$ is associated with the probability amplitude of creating a Stokes--anti-Stokes pair in the medium given that a pair of laser photons with frequencies $\omega_\ell$ and $\omega_\ell'$ propagates in it.
In order to see that, we write schematically $P^{(0)} \propto a_\ell b + a_\ell b^\dagger$ and $P^{(1)} \propto \chi^{(3)} a_\ell a_\ell' a_S^\dagger$, where $a_\ell$ and $a_\ell'$ are laser photon annihilation operators, $b$ is the analogous for the phonon, and $a_S$ and $a_A$ for Stokes and anti-Stokes photons.
We write the corrections in the interaction energy $H_I \propto E_i P_i$, due to $P^{(0)}$ and $P^{(1)}$, respectively,
\begin{subequations}
    \begin{equation}
    H_I^{(0)} \propto a_\ell b a_A^\dagger + a_\ell b^\dagger a_S^\dagger
    + \text{H.c.},
    \end{equation}
    \begin{equation}
    H_I^{(1)} \propto \chi^{(3)} a_\ell a_\ell' a_S^\dagger a_A^\dagger
    + \text{H.c.},
    \end{equation}
\end{subequations}
where H.c. represents the Hermitian conjugate and we only represented the terms relevant for us, to avoid cumbersomeness.
It is clear the $P^{(0)}$ contains one interaction in it for the creation of one Stokes or anti-Stokes photon,
while $P^{(1)}$ contains two Raman interactions, as can be attested by the two $b$ operators in the commutator in $\chi^{(3)} \propto [b,b^\dagger]$ (see (\ref{eq:pol_first})).
When $H_I^{(1)}$ is applied to an initially coherent state in the laser modes, with amplitudes $v_\ell$ and $v_\ell'$, and vacuum of SaS modes,
\begin{equation}\label{eq:SaS_state}
    H_I^{(1)} |v_\ell, v_\ell', 0_S, 0_A \rangle
    \propto
    \chi^{(3)} v_\ell v_\ell' |v_\ell, v_\ell', 1_S, 1_A \rangle,
\end{equation}
we see that the probability of the transition is proportional to $|\chi^{(3)}|^2$ and the phase of the created two-photon state is proportional to $\arg(\chi^{(3)})$.
On resonance, $\omega_\ell'-\omega_S = \omegatil$, the susceptibility $\chi^{(3)}$ becomes purely imaginary, but care must be taken in its interpretation because it is a non-linear susceptibility.
In this case, the imaginary part of the susceptibility does not correspond to absorption, but can be related to a phase shift of the generated fields \cite{bloembergen}, as seen in Equation (\ref{eq:SaS_state}).

Note that $(H_I^{(0)})^2$ is of the same order as $H_I^{(1)}$, as both contain two interactions, so $P^{(0)}$ may also contribute to the formation of SaS photon pairs, though not in the FWM form of Equation (\ref{eq:pol_fAS}).
In particular, $P^{(0)}$ creates real phonons, so it cannot create Stokes or anti-Stokes photons out of the Raman resonance, but it can have a contribution to the resonant SaS pairs \cite{junior2020}.
It can also create independent Stokes and anti-Stokes photons that do not carry correlations, which can hide the quantum correlations in the resonant SaS pairs \cite{saraiva2017,freitas2024,sier2024}.
We are currently working on the theoretical characterization of this contribution, which has the important feature of being sensitive to the material temperature, but should only be relevant on the Raman resonance.
The $\chi^{(3)}$ derived from $P^{(1)}$ describes only the contribution in which no real phonons are created in the process.
When away from the Raman resonance, this is the only contribution, so the quantum state of the SaS photon pairs can be accurately described by it \cite{sier2024}.

In principle, one could go further on the perturbative framework, calculating higher orders of perturbation,
allowing for more and more interactions in any one scattering event and going to higher non-linearities (e.g. six-photon interaction).
However, Raman scattering is so weak that anything beyond SaS pair generation is extremely unlikely and can be neglected.

\begin{figure}[ht]
\begin{center}
    \includegraphics[width=.65\textwidth]{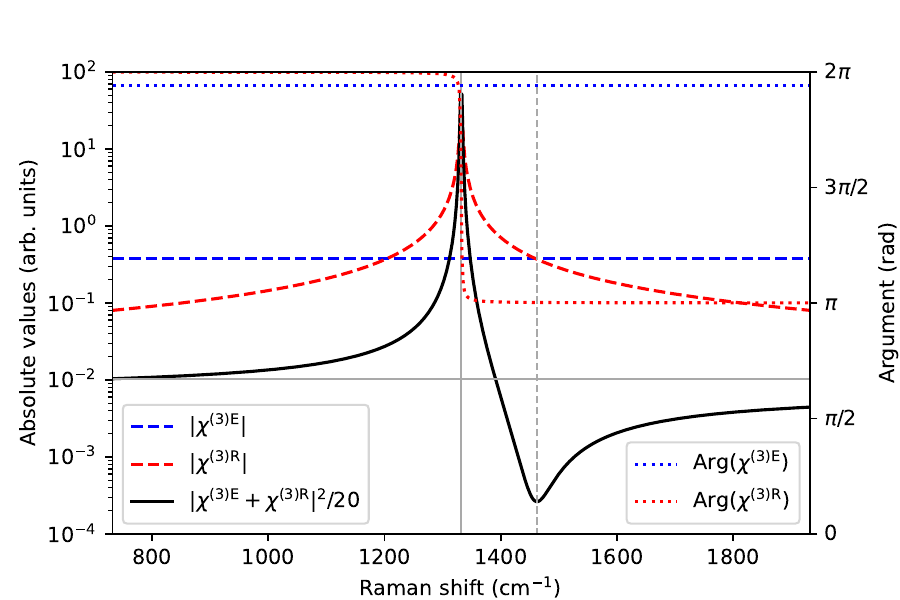}
\end{center}
    \caption{\label{fig:chi_arg} Sum of vibrational (superscript R, red lines) and electronic (superscript E, blue lines) third-order susceptibilities of the SaS scattering, $\chi^{(3)\T{R}}$ and $\chi^{(3)\T{E}}$, in terms of the Raman shift $(\omega_\ell' - \omega_S)$, according to (\ref{eq:chi3_R}), (\ref{eq:chi3_E}) and (\ref{eq:chi3_sum}).
        The absolute parts of $\chi^{(3)\T{R}}$ and $\chi^{(3)\T{E}}$ are dashed lines, and the arguments are dotted lines.
        In the plots, we consider the diamond values $\omegatil = 1332 \text{ cm}^{-1}$, $\gamma = (3 \text{ ps})^{-1}$, and $A^{\T{R}}_{ijj'i'} = 171 $ arb. units and $A^{\T{E}}_{ijj'i'} = (0.37 - 0.07 i)$ arb. units, the same values used to fit the experiment of \cite{sier2024}.
        The resulting total susceptibility square modulus $|\chi^{(3)\T{R}}+\chi^{(3)\T{E}}|^2$ is plotted as a black solid line, rescaled by a factor of 1/20 for legibility.
        The solid vertical grey line marks the vibrational Raman resonance,
        the dashed vertical grey line marks the minimum of the susceptibility, where $|\chi^{(3)\T{R}}|=|\chi^{(3)\T{E}}|$ in the region above the resonance,
        and the solid horizontal grey line is a guide to the eye, to stress that the susceptibility is lower above than below resonance.
    }
\end{figure}

At this point, one has to remember that vibrational Raman is not the whole picture in the FWM process of the SaS scattering.
In fact, since the electrons mediate the interaction, there will be a FWM contribution due to the interaction of light directly with the electronic degrees of freedom.
If the electronic gap is much higher than the laser energy (which is true for diamond and silicon, for instance), the electronic FWM will always be non-resonant in the region of the vibrational Raman resonance \cite{bloembergen, levenson72}.
We can then approximate the susceptibility of the electronic FWM to a constant with an energy conservation condition,
\begin{equation}\label{eq:chi3_E}
    \chi^{(3) \T{E}}_{ijj'i'} (-\omega_A, \omega_\ell, \omega_\ell', -\omega_S)
    =
    A^{\T{E}}_{ijj'i'}
    \delta( \omega_{\ell} + \omega_{\ell}' - \omega_{S} -\omega_A).
\end{equation}
In Figure \ref{fig:chi_arg}, we plot what happens when the two susceptibilities are summed up, such that the total susceptibility of the process is
\begin{equation}
    \chi^{(3)}_{ijj'i'} (-\omega_A, \omega_\ell, \omega_\ell', -\omega_S)
    =
    \chi^{(3) \T{R}}_{ijj'i'} (-\omega_A, \omega_\ell, \omega_\ell', -\omega_S)
    +
    \chi^{(3) \T{E}}_{ijj'i'} (-\omega_A, \omega_\ell, \omega_\ell', -\omega_S),
\end{equation}
or more explicitly
\begin{equation}\label{eq:chi3_sum}
    \chi^{(3)}_{ijj'i'} (-\omega_A, \omega_\ell, \omega_\ell', -\omega_S)
    =
    \left[
        A^{\T{R}}_{ijj'i'}
        \frac{\gamma}
        {( \omega_{\ell} - \omega_{S} - \omegatil + i \gamma/2
        )}
        +
        A^{\T{E}}_{ijj'i'}
    \right]
    \delta( \omega_{\ell} + \omega_{\ell}' - \omega_{S} -\omega_A)
    .
\end{equation}
The phonon resonance introduces a $\pi$ phase shift in $\chi^{(3) \T{R}}$ with respect to the Raman shift $(\omega_\ell'-\omega_S)$ at $\omegatil$, because the third-order FWM susceptibility is proportional to the probability amplitude of the phonon transitions.
The electronic susceptibility, however, is too far from resonance, so when one crosses $\omegatil$, the sum of the two susceptibilities add up below $\omegatil$, but subtract above $\omegatil$.
Quantum mechanically, this is associated with the fact that the two FWM scatterings, vibrational and electronic, produce the same state output from the same state input, and thus their probability amplitudes must be added coherently.
This is an interference effect between the two kinds of transitions, which provides a microscopic explanation of the previously reported asymmetry in the SaS spectrum, in which the probability of generating pairs above resonance is lower than below it \cite{junior2019, sier2024}.

The validity of our approach has been tested against experiment with very good agreement \cite{sier2024}, and a similar theoretical framework has been independently used elsewhere, also with good experimental agreement \cite{vento2025}.
As discussed previously, the FWM contribution from $P^{(1)}$ can be used alone to calculate the correlated SaS photon pair spectrum out of resonance, while $P^{(0)}$ also contributes on resonance and can be used to calculate independent Stokes and anti-Stokes production rate.
With these two quantities, one is able to model the correlation $g^{(2)}(0)$, which characterizes non-classical correlations in the SaS pair \cite{saraiva2017, sier2024}.
The overall shape of the SaS spectrum depends on the relative amplitude between the resonant (R) and the flat (E) term, which is given by the tensorial components of the susceptibilities, $A^{\T{R}}_{ijj'i'}$ and $A^{\T{E}}_{ijj'i'}$,
and the good spectral agreement can be used to obtain the values of the tensor components.
Furthermore, the simultaneous consideration of the vibrational Raman and electronic FWM susceptibilities is used to explain the polarization entanglement in the SaS photon pairs.
It is shown that by changing the crystal orientation with respect to the excitation laser polarization yields different $A^{\T{R}}_{ijj'i'}$ and $A^{\T{E}}_{ijj'i'}$ to be combined in the polarization of the scattered photons, and this can be exploited to tune the frequency of maximally entangled SaS photon pairs \cite{sier2024}.

\section{Conclusion}\label{sec:conclusion}

In conclusion, we have presented a fully quantum theory for the spectral properties of the FWM contribution to the correlated SaS scattering.
This theory, here developed from first principles, was missing from the literature on the quantum properties of the SaS photon pair generation, developed in the last decade.
We can, with it, explain the asymmetry in the correlated SaS spectrum that appeared in previous experiments and lacked a proper explanation \cite{junior2019, sier2024}.
In addition, it becomes clear that the material quantum state cannot be probed by the purely FWM interaction, which contributes to the SaS scattering with the quantum correlations,
differently from simple Raman, in which the temperature of the material has a role in the scattering probability of both Stokes and anti-Stokes modes.
We have also shown that the susceptibility derived in the stimulated Raman context \cite{bloembergen, shen, levenson74}, involving classical intense fields, has the same spectral shape as the spontaneous SaS pair generation, in which the fields are quantized, providing a robust theoretical ground for the assumptions of recent experiments \cite{sier2024, vento2025}.

Our formalism, using Heisenberg operators, can be applied to a variety of initial quantum states, and not only to the vacuum.
This means that, beyond the generation of photon pairs, we can for instance calculate the quantum effects of the SaS scattering on Stokes or anti-Stokes beams initially in a weak coherent state, that acts as a stimulating mode and will transition to photon-added coherent states.
We thus open the path to the study of intermediate regimes between the classical stimulated Raman and the spontaneous generation of SaS photon pairs.

\begin{acknowledgments}
This work was supported by the IDOR/Pioneer Science Initiative (www.pioneerscience.org),
FAPEMIG -- Funda\c c\~ao de Amparo \`a Pesquisa do Estado de Minas Gerais, Project APQ-04852-23 --
and CNPq -- Conselho Nacional de Desenvolvimento Cient\'ifico e Tecnol\'ogico, Project 421469/2023-4.
M.F.S. acknowledges support from the Brazilian National Institute of Science and Technology for Quantum Information [CNPq INCT-IQ 465469/2014-0],
FAPERJ -- Funda\c c\~ao Carlos Chagas Filho de Amparo \`a Pesquisa no Estado do Rio de Janeiro, Projects CNE E-26/200.307/2023 and E-26/210.069/2020 --
and CNPq -- Conselho Nacional de Desenvolvimento Cient\'ifico e Tecnol\'ogico, Project 302872/2019-1.
\end{acknowledgments}


\begin{thebibliography}{21}%
\makeatletter
\providecommand \@ifxundefined [1]{%
 \@ifx{#1\undefined}
}%
\providecommand \@ifnum [1]{%
 \ifnum #1\expandafter \@firstoftwo
 \else \expandafter \@secondoftwo
 \fi
}%
\providecommand \@ifx [1]{%
 \ifx #1\expandafter \@firstoftwo
 \else \expandafter \@secondoftwo
 \fi
}%
\providecommand \natexlab [1]{#1}%
\providecommand \enquote  [1]{``#1''}%
\providecommand \bibnamefont  [1]{#1}%
\providecommand \bibfnamefont [1]{#1}%
\providecommand \citenamefont [1]{#1}%
\providecommand \href@noop [0]{\@secondoftwo}%
\providecommand \href [0]{\begingroup \@sanitize@url \@href}%
\providecommand \@href[1]{\@@startlink{#1}\@@href}%
\providecommand \@@href[1]{\endgroup#1\@@endlink}%
\providecommand \@sanitize@url [0]{\catcode `\\12\catcode `\$12\catcode
  `\&12\catcode `\#12\catcode `\^12\catcode `\_12\catcode `\%12\relax}%
\providecommand \@@startlink[1]{}%
\providecommand \@@endlink[0]{}%
\providecommand \url  [0]{\begingroup\@sanitize@url \@url }%
\providecommand \@url [1]{\endgroup\@href {#1}{\urlprefix }}%
\providecommand \urlprefix  [0]{URL }%
\providecommand \Eprint [0]{\href }%
\providecommand \doibase [0]{https://doi.org/}%
\providecommand \selectlanguage [0]{\@gobble}%
\providecommand \bibinfo  [0]{\@secondoftwo}%
\providecommand \bibfield  [0]{\@secondoftwo}%
\providecommand \translation [1]{[#1]}%
\providecommand \BibitemOpen [0]{}%
\providecommand \bibitemStop [0]{}%
\providecommand \bibitemNoStop [0]{.\EOS\space}%
\providecommand \EOS [0]{\spacefactor3000\relax}%
\providecommand \BibitemShut  [1]{\csname bibitem#1\endcsname}%
\let\auto@bib@innerbib\@empty
\bibitem [{\citenamefont {Bloembergen}(1996)}]{bloembergen}%
  \BibitemOpen
  \bibfield  {author} {\bibinfo {author} {\bibfnamefont {N.}~\bibnamefont
  {Bloembergen}},\ }\href@noop {} {\emph {\bibinfo {title} {Nonlinear
  Optics}}}\ (\bibinfo  {publisher} {World Scientific Publishing Company},\
  \bibinfo {address} {New York},\ \bibinfo {year} {1996})\BibitemShut {NoStop}%
\bibitem [{\citenamefont {von Foerster}\ and\ \citenamefont
  {Glauber}(1971)}]{foerster71}%
  \BibitemOpen
  \bibfield  {author} {\bibinfo {author} {\bibfnamefont {T.}~\bibnamefont {von
  Foerster}}\ and\ \bibinfo {author} {\bibfnamefont {R.~J.}\ \bibnamefont
  {Glauber}},\ }\bibfield  {title} {\bibinfo {title} {Quantum theory of light
  propagation in amplifying media},\ }\href
  {https://doi.org/10.1103/PhysRevA.3.1484} {\bibfield  {journal} {\bibinfo
  {journal} {Phys. Rev. A}\ }\textbf {\bibinfo {volume} {3}},\ \bibinfo {pages}
  {1484} (\bibinfo {year} {1971})}\BibitemShut {NoStop}%
\bibitem [{\citenamefont {Shen}(1984)}]{shen}%
  \BibitemOpen
  \bibfield  {author} {\bibinfo {author} {\bibfnamefont {Y.~R.}\ \bibnamefont
  {Shen}},\ }\href@noop {} {\emph {\bibinfo {title} {The principles of
  Nonlinear Optics}}}\ (\bibinfo  {publisher} {Wiley-Interscience},\ \bibinfo
  {address} {New Jersey},\ \bibinfo {year} {1984})\BibitemShut {NoStop}%
\bibitem [{\citenamefont {Boyd}(2020)}]{boyd}%
  \BibitemOpen
  \bibfield  {author} {\bibinfo {author} {\bibfnamefont {R.~W.}\ \bibnamefont
  {Boyd}},\ }\href@noop {} {\emph {\bibinfo {title} {Nonlinear optics}}},\
  \bibinfo {edition} {4th}\ ed.\ (\bibinfo  {publisher} {Academic Press},\
  \bibinfo {address} {London},\ \bibinfo {year} {2020})\BibitemShut {NoStop}%
\bibitem [{\citenamefont {Klyshko}(1977)}]{klyshko1977}%
  \BibitemOpen
  \bibfield  {author} {\bibinfo {author} {\bibfnamefont {D.~N.}\ \bibnamefont
  {Klyshko}},\ }\bibfield  {title} {\bibinfo {title} {Correlation between the
  {S}tokes and anti-{S}tokes components in inelastic scattering of light},\
  }\href {https://doi.org/10.1070/qe1977v007n06abeh012890} {\bibfield
  {journal} {\bibinfo  {journal} {Soviet Journal of Quantum Electronics}\
  }\textbf {\bibinfo {volume} {7}},\ \bibinfo {pages} {755} (\bibinfo {year}
  {1977})}\BibitemShut {NoStop}%
\bibitem [{\citenamefont {Kasperczyk}\ \emph {et~al.}(2015)\citenamefont
  {Kasperczyk}, \citenamefont {Jorio}, \citenamefont {Neu}, \citenamefont
  {Maletinsky},\ and\ \citenamefont {Novotny}}]{kasperczyk2015}%
  \BibitemOpen
  \bibfield  {author} {\bibinfo {author} {\bibfnamefont {M.}~\bibnamefont
  {Kasperczyk}}, \bibinfo {author} {\bibfnamefont {A.}~\bibnamefont {Jorio}},
  \bibinfo {author} {\bibfnamefont {E.}~\bibnamefont {Neu}}, \bibinfo {author}
  {\bibfnamefont {P.}~\bibnamefont {Maletinsky}},\ and\ \bibinfo {author}
  {\bibfnamefont {L.}~\bibnamefont {Novotny}},\ }\bibfield  {title} {\bibinfo
  {title} {Stokes--anti-{S}tokes correlations in diamond},\ }\href
  {https://doi.org/10.1364/OL.40.002393} {\bibfield  {journal} {\bibinfo
  {journal} {Opt. Lett.}\ }\textbf {\bibinfo {volume} {40}},\ \bibinfo {pages}
  {2393} (\bibinfo {year} {2015})}\BibitemShut {NoStop}%
\bibitem [{\citenamefont {Saraiva}\ \emph {et~al.}(2017)\citenamefont
  {Saraiva}, \citenamefont {J\'unior}, \citenamefont {de~Melo~e Souza},
  \citenamefont {Pena}, \citenamefont {Monken}, \citenamefont {Santos},
  \citenamefont {Koiller},\ and\ \citenamefont {Jorio}}]{saraiva2017}%
  \BibitemOpen
  \bibfield  {author} {\bibinfo {author} {\bibfnamefont {A.}~\bibnamefont
  {Saraiva}}, \bibinfo {author} {\bibfnamefont {F.~S. d.~A.}\ \bibnamefont
  {J\'unior}}, \bibinfo {author} {\bibfnamefont {R.}~\bibnamefont {de~Melo~e
  Souza}}, \bibinfo {author} {\bibfnamefont {A.~P.}\ \bibnamefont {Pena}},
  \bibinfo {author} {\bibfnamefont {C.~H.}\ \bibnamefont {Monken}}, \bibinfo
  {author} {\bibfnamefont {M.~F.}\ \bibnamefont {Santos}}, \bibinfo {author}
  {\bibfnamefont {B.}~\bibnamefont {Koiller}},\ and\ \bibinfo {author}
  {\bibfnamefont {A.}~\bibnamefont {Jorio}},\ }\bibfield  {title} {\bibinfo
  {title} {Photonic counterparts of {C}ooper pairs},\ }\href
  {https://doi.org/10.1103/PhysRevLett.119.193603} {\bibfield  {journal}
  {\bibinfo  {journal} {Phys. Rev. Lett.}\ }\textbf {\bibinfo {volume} {119}},\
  \bibinfo {pages} {193603} (\bibinfo {year} {2017})}\BibitemShut {NoStop}%
\bibitem [{\citenamefont {J\'unior}\ \emph {et~al.}(2019)\citenamefont
  {J\'unior}, \citenamefont {Saraiva}, \citenamefont {Santos}, \citenamefont
  {Koiller}, \citenamefont {Souza}, \citenamefont {Pena}, \citenamefont
  {Silva}, \citenamefont {Monken},\ and\ \citenamefont {Jorio}}]{junior2019}%
  \BibitemOpen
  \bibfield  {author} {\bibinfo {author} {\bibfnamefont {F.~S. d.~A.}\
  \bibnamefont {J\'unior}}, \bibinfo {author} {\bibfnamefont {A.}~\bibnamefont
  {Saraiva}}, \bibinfo {author} {\bibfnamefont {M.~F.}\ \bibnamefont {Santos}},
  \bibinfo {author} {\bibfnamefont {B.}~\bibnamefont {Koiller}}, \bibinfo
  {author} {\bibfnamefont {R.~d. M.~e.}\ \bibnamefont {Souza}}, \bibinfo
  {author} {\bibfnamefont {A.~P.}\ \bibnamefont {Pena}}, \bibinfo {author}
  {\bibfnamefont {R.~A.}\ \bibnamefont {Silva}}, \bibinfo {author}
  {\bibfnamefont {C.~H.}\ \bibnamefont {Monken}},\ and\ \bibinfo {author}
  {\bibfnamefont {A.}~\bibnamefont {Jorio}},\ }\bibfield  {title} {\bibinfo
  {title} {Stokes--anti-{S}tokes correlated photon properties akin to photonic
  cooper pairs},\ }\href {https://doi.org/10.1103/PhysRevB.99.100503}
  {\bibfield  {journal} {\bibinfo  {journal} {Phys. Rev. B}\ }\textbf {\bibinfo
  {volume} {99}},\ \bibinfo {pages} {100503} (\bibinfo {year}
  {2019})}\BibitemShut {NoStop}%
\bibitem [{\citenamefont {J\'unior}\ \emph {et~al.}(2020)\citenamefont
  {J\'unior}, \citenamefont {Santos}, \citenamefont {Monken},\ and\
  \citenamefont {Jorio}}]{junior2020}%
  \BibitemOpen
  \bibfield  {author} {\bibinfo {author} {\bibfnamefont {F.~S. d.~A.}\
  \bibnamefont {J\'unior}}, \bibinfo {author} {\bibfnamefont {M.~F.}\
  \bibnamefont {Santos}}, \bibinfo {author} {\bibfnamefont {C.~H.}\
  \bibnamefont {Monken}},\ and\ \bibinfo {author} {\bibfnamefont
  {A.}~\bibnamefont {Jorio}},\ }\bibfield  {title} {\bibinfo {title} {Lifetime
  and polarization for real and virtual correlated stokes-anti-stokes raman
  scattering in diamond},\ }\href
  {https://doi.org/10.1103/PhysRevResearch.2.013084} {\bibfield  {journal}
  {\bibinfo  {journal} {Phys. Rev. Research}\ }\textbf {\bibinfo {volume}
  {2}},\ \bibinfo {pages} {013084} (\bibinfo {year} {2020})}\BibitemShut
  {NoStop}%
\bibitem [{\citenamefont {Freitas}\ \emph {et~al.}(2023)\citenamefont
  {Freitas}, \citenamefont {Machado}, \citenamefont {Valente}, \citenamefont
  {Sier}, \citenamefont {Corr\^ea}, \citenamefont {Saito}, \citenamefont
  {Galland}, \citenamefont {Santos}, \citenamefont {Monken},\ and\
  \citenamefont {Jorio}}]{freitas2023}%
  \BibitemOpen
  \bibfield  {author} {\bibinfo {author} {\bibfnamefont {T.~A.}\ \bibnamefont
  {Freitas}}, \bibinfo {author} {\bibfnamefont {P.}~\bibnamefont {Machado}},
  \bibinfo {author} {\bibfnamefont {L.}~\bibnamefont {Valente}}, \bibinfo
  {author} {\bibfnamefont {D.}~\bibnamefont {Sier}}, \bibinfo {author}
  {\bibfnamefont {R.}~\bibnamefont {Corr\^ea}}, \bibinfo {author}
  {\bibfnamefont {R.}~\bibnamefont {Saito}}, \bibinfo {author} {\bibfnamefont
  {C.}~\bibnamefont {Galland}}, \bibinfo {author} {\bibfnamefont {M.~F.}\
  \bibnamefont {Santos}}, \bibinfo {author} {\bibfnamefont {C.~H.}\
  \bibnamefont {Monken}},\ and\ \bibinfo {author} {\bibfnamefont
  {A.}~\bibnamefont {Jorio}},\ }\bibfield  {title} {\bibinfo {title}
  {Microscopic origin of polarization-entangled stokes--anti-stokes photons in
  diamond},\ }\href {https://doi.org/10.1103/PhysRevA.108.L051501} {\bibfield
  {journal} {\bibinfo  {journal} {Phys. Rev. A}\ }\textbf {\bibinfo {volume}
  {108}},\ \bibinfo {pages} {L051501} (\bibinfo {year} {2023})}\BibitemShut
  {NoStop}%
\bibitem [{\citenamefont {Freitas}\ \emph {et~al.}(2024)\citenamefont
  {Freitas}, \citenamefont {Machado}, \citenamefont {Saito}, \citenamefont
  {Santos}, \citenamefont {Monken},\ and\ \citenamefont {Jorio}}]{freitas2024}%
  \BibitemOpen
  \bibfield  {author} {\bibinfo {author} {\bibfnamefont {T.~A.}\ \bibnamefont
  {Freitas}}, \bibinfo {author} {\bibfnamefont {P.}~\bibnamefont {Machado}},
  \bibinfo {author} {\bibfnamefont {R.}~\bibnamefont {Saito}}, \bibinfo
  {author} {\bibfnamefont {M.~F.}\ \bibnamefont {Santos}}, \bibinfo {author}
  {\bibfnamefont {C.~H.}\ \bibnamefont {Monken}},\ and\ \bibinfo {author}
  {\bibfnamefont {A.}~\bibnamefont {Jorio}},\ }\bibfield  {title} {\bibinfo
  {title} {Polarization state tomography of {S}tokes--anti-{S}tokes photon
  pairs from degenerate four-wave mixing in diamond},\ }\href
  {https://doi.org/10.1103/PhysRevA.110.053714} {\bibfield  {journal} {\bibinfo
   {journal} {Phys. Rev. A}\ }\textbf {\bibinfo {volume} {110}},\ \bibinfo
  {pages} {053714} (\bibinfo {year} {2024})}\BibitemShut {NoStop}%
\bibitem [{\citenamefont {Freitas}\ \emph {et~al.}(2025)\citenamefont
  {Freitas}, \citenamefont {Machado}, \citenamefont {Santos}, \citenamefont
  {Monken},\ and\ \citenamefont {Jorio}}]{freitas2025}%
  \BibitemOpen
  \bibfield  {author} {\bibinfo {author} {\bibfnamefont {T.~A.}\ \bibnamefont
  {Freitas}}, \bibinfo {author} {\bibfnamefont {P.}~\bibnamefont {Machado}},
  \bibinfo {author} {\bibfnamefont {M.~F.}\ \bibnamefont {Santos}}, \bibinfo
  {author} {\bibfnamefont {C.~H.}\ \bibnamefont {Monken}},\ and\ \bibinfo
  {author} {\bibfnamefont {A.}~\bibnamefont {Jorio}},\ }\bibfield  {title}
  {\bibinfo {title} {Characterization of the polarization state of correlated
  {S}tokes–anti-{S}tokes {R}aman scattering in diamond},\ }\href
  {https://doi.org/10.1002/pssb.202400275} {\bibfield  {journal} {\bibinfo
  {journal} {Phys. Status Solidi B}\ }\textbf {\bibinfo {volume} {262}},\
  \bibinfo {pages} {2400275} (\bibinfo {year} {2025})},\ \Eprint
  {https://arxiv.org/abs/https://onlinelibrary.wiley.com/doi/pdf/10.1002/pssb.202400275}
  {https://onlinelibrary.wiley.com/doi/pdf/10.1002/pssb.202400275} \BibitemShut
  {NoStop}%
\bibitem [{\citenamefont {Vento}\ \emph {et~al.}(2025)\citenamefont {Vento},
  \citenamefont {Ciccarello}, \citenamefont {Amirtharaj},\ and\ \citenamefont
  {Galland}}]{vento2025}%
  \BibitemOpen
  \bibfield  {author} {\bibinfo {author} {\bibfnamefont {V.}~\bibnamefont
  {Vento}}, \bibinfo {author} {\bibfnamefont {F.}~\bibnamefont {Ciccarello}},
  \bibinfo {author} {\bibfnamefont {S.~P.}\ \bibnamefont {Amirtharaj}},\ and\
  \bibinfo {author} {\bibfnamefont {C.}~\bibnamefont {Galland}},\ }\bibfield
  {title} {\bibinfo {title} {How to use the dispersion in the
  ${\ensuremath{\chi}}^{(3)}$ tensor for broadband generation of
  polarization-entangled photons},\ }\href
  {https://doi.org/10.1103/PhysRevResearch.7.L022017} {\bibfield  {journal}
  {\bibinfo  {journal} {Phys. Rev. Res.}\ }\textbf {\bibinfo {volume} {7}},\
  \bibinfo {pages} {L022017} (\bibinfo {year} {2025})}\BibitemShut {NoStop}%
\bibitem [{\citenamefont {Timsina}\ \emph {et~al.}(2024)\citenamefont
  {Timsina}, \citenamefont {Hammadia}, \citenamefont {Milani}, \citenamefont
  {J\'unior}, \citenamefont {Brolo},\ and\ \citenamefont
  {de~Sousa}}]{timsina2024}%
  \BibitemOpen
  \bibfield  {author} {\bibinfo {author} {\bibfnamefont {S.}~\bibnamefont
  {Timsina}}, \bibinfo {author} {\bibfnamefont {T.}~\bibnamefont {Hammadia}},
  \bibinfo {author} {\bibfnamefont {S.~G.}\ \bibnamefont {Milani}}, \bibinfo
  {author} {\bibfnamefont {F.~S. d.~A.}\ \bibnamefont {J\'unior}}, \bibinfo
  {author} {\bibfnamefont {A.}~\bibnamefont {Brolo}},\ and\ \bibinfo {author}
  {\bibfnamefont {R.}~\bibnamefont {de~Sousa}},\ }\bibfield  {title} {\bibinfo
  {title} {Resonant squeezed light from photonic {C}ooper pairs},\ }\href
  {https://doi.org/10.1103/PhysRevResearch.6.033067} {\bibfield  {journal}
  {\bibinfo  {journal} {Phys. Rev. Res.}\ }\textbf {\bibinfo {volume} {6}},\
  \bibinfo {pages} {033067} (\bibinfo {year} {2024})}\BibitemShut {NoStop}%
\bibitem [{\citenamefont {Louisell}(1973)}]{louisell}%
  \BibitemOpen
  \bibfield  {author} {\bibinfo {author} {\bibfnamefont {W.~H.}\ \bibnamefont
  {Louisell}},\ }\href@noop {} {\emph {\bibinfo {title} {Quantum statistical
  properties of radiation}}}\ (\bibinfo  {publisher} {Wiley},\ \bibinfo
  {address} {New York},\ \bibinfo {year} {1973})\BibitemShut {NoStop}%
\bibitem [{\citenamefont {Levenson}\ and\ \citenamefont
  {Bloembergen}(1974)}]{levenson74}%
  \BibitemOpen
  \bibfield  {author} {\bibinfo {author} {\bibfnamefont {M.~D.}\ \bibnamefont
  {Levenson}}\ and\ \bibinfo {author} {\bibfnamefont {N.}~\bibnamefont
  {Bloembergen}},\ }\bibfield  {title} {\bibinfo {title} {Dispersion of the
  nonlinear optical susceptibility tensor in centrosymmetric media},\ }\href
  {https://doi.org/10.1103/PhysRevB.10.4447} {\bibfield  {journal} {\bibinfo
  {journal} {Phys. Rev. B}\ }\textbf {\bibinfo {volume} {10}},\ \bibinfo
  {pages} {4447} (\bibinfo {year} {1974})}\BibitemShut {NoStop}%
\bibitem [{\citenamefont {Raymer}(2020)}]{raymer2020}%
  \BibitemOpen
  \bibfield  {author} {\bibinfo {author} {\bibfnamefont {M.~G.}\ \bibnamefont
  {Raymer}},\ }\bibfield  {title} {\bibinfo {title} {Quantum theory of light in
  a dispersive structured linear dielectric: a macroscopic {H}amiltonian
  tutorial treatment},\ }\href {https://doi.org/10.1080/09500340.2019.1706773}
  {\bibfield  {journal} {\bibinfo  {journal} {J. Mod. Opt.}\ }\textbf {\bibinfo
  {volume} {67}},\ \bibinfo {pages} {196} (\bibinfo {year} {2020})}\BibitemShut
  {NoStop}%
\bibitem [{\citenamefont {van Kampen}(2008)}]{vankampen}%
  \BibitemOpen
  \bibfield  {author} {\bibinfo {author} {\bibfnamefont {N.~G.}\ \bibnamefont
  {van Kampen}},\ }\href@noop {} {\emph {\bibinfo {title} {Stochastic processes
  in physics and chemistry}}}\ (\bibinfo  {publisher} {Elsevier},\ \bibinfo
  {address} {Amsterdam},\ \bibinfo {year} {2008})\BibitemShut {NoStop}%
\bibitem [{\citenamefont {Guimar\~aes}\ \emph {et~al.}(2020)\citenamefont
  {Guimar\~aes}, \citenamefont {Santos}, \citenamefont {Jorio},\ and\
  \citenamefont {Monken}}]{guimaraes2020}%
  \BibitemOpen
  \bibfield  {author} {\bibinfo {author} {\bibfnamefont {A.~V.~A.}\
  \bibnamefont {Guimar\~aes}}, \bibinfo {author} {\bibfnamefont {M.~F.}\
  \bibnamefont {Santos}}, \bibinfo {author} {\bibfnamefont {A.}~\bibnamefont
  {Jorio}},\ and\ \bibinfo {author} {\bibfnamefont {C.~H.}\ \bibnamefont
  {Monken}},\ }\bibfield  {title} {\bibinfo {title} {Stokes--anti-{S}tokes
  light-scattering process: A photon-wave-function approach},\ }\href
  {https://doi.org/10.1103/PhysRevA.102.033719} {\bibfield  {journal} {\bibinfo
   {journal} {Phys. Rev. A}\ }\textbf {\bibinfo {volume} {102}},\ \bibinfo
  {pages} {033719} (\bibinfo {year} {2020})}\BibitemShut {NoStop}%
\bibitem [{\citenamefont {Sier}\ \emph {et~al.}(2025)\citenamefont {Sier},
  \citenamefont {Valente}, \citenamefont {Freitas}, \citenamefont {Santos},
  \citenamefont {Monken}, \citenamefont {Corr\^ea},\ and\ \citenamefont
  {Jorio}}]{sier2024}%
  \BibitemOpen
  \bibfield  {author} {\bibinfo {author} {\bibfnamefont {D.}~\bibnamefont
  {Sier}}, \bibinfo {author} {\bibfnamefont {L.}~\bibnamefont {Valente}},
  \bibinfo {author} {\bibfnamefont {T.~A.}\ \bibnamefont {Freitas}}, \bibinfo
  {author} {\bibfnamefont {M.~F.}\ \bibnamefont {Santos}}, \bibinfo {author}
  {\bibfnamefont {C.~H.}\ \bibnamefont {Monken}}, \bibinfo {author}
  {\bibfnamefont {R.}~\bibnamefont {Corr\^ea}},\ and\ \bibinfo {author}
  {\bibfnamefont {A.}~\bibnamefont {Jorio}},\ }\bibfield  {title} {\bibinfo
  {title} {Emergent broadband polarization entanglement from electronic and
  phononic stokes--anti-stokes indistinguishability},\ }\href
  {https://doi.org/10.1103/c8nj-chfn} {\bibfield  {journal} {\bibinfo
  {journal} {Phys. Rev. A}\ }\textbf {\bibinfo {volume} {112}},\ \bibinfo
  {pages} {033702} (\bibinfo {year} {2025})}\BibitemShut {NoStop}%
\bibitem [{\citenamefont {Levenson}\ \emph {et~al.}(1972)\citenamefont
  {Levenson}, \citenamefont {Flytzanis},\ and\ \citenamefont
  {Bloembergen}}]{levenson72}%
  \BibitemOpen
  \bibfield  {author} {\bibinfo {author} {\bibfnamefont {M.~D.}\ \bibnamefont
  {Levenson}}, \bibinfo {author} {\bibfnamefont {C.}~\bibnamefont
  {Flytzanis}},\ and\ \bibinfo {author} {\bibfnamefont {N.}~\bibnamefont
  {Bloembergen}},\ }\bibfield  {title} {\bibinfo {title} {Interference of
  resonant and nonresonant three-wave mixing in diamond},\ }\href
  {https://doi.org/10.1103/PhysRevB.6.3962} {\bibfield  {journal} {\bibinfo
  {journal} {Phys. Rev. B}\ }\textbf {\bibinfo {volume} {6}},\ \bibinfo {pages}
  {3962} (\bibinfo {year} {1972})}\BibitemShut {NoStop}%
\end{thebibliography}

%

\appendix

\section{Commutator of phonon operators}\label{app:commutator}

In this section we calculate the two-time commutator of phonon bosonic operators $[ \bar{b}_{\vec{q} \eta} (t), \bar{b}_{\vec{q}' \eta'}^\dagger (t-t')]$, which is obtained from the solution $\bar{b}_{\vec{q}\eta}(t)$, Eq. (\ref{eq:phonon_free_solution}) of the main article,
\begin{eqnarray}
    [ \bar{b}_{\vec{q} \eta} (t), \bar{b}_{\vec{q}' \eta'}^\dagger (t-t')]
    &=&
    [ b_{\vec{q} \eta} (t_0), b_{\vec{q}' \eta'}^\dagger (t_0) ]
        e^{-i( \omegatil_{\eta} -i \gamma_\eta/2 ) (t-t_0)}
        e^{i( \omegatil_{\eta'}' +i \gamma_{\eta'}/2 ) (t-t'-t_0)}
    \notag\\
    && +
    [ C_{\vec{q} \eta} (t_0), C_{\vec{q}' \eta'}^\dagger (t_0) ]
        e^{-i( \omegatil_{\eta} -i \gamma_\eta/2 ) (t-t_0)}
        e^{i( \omegatil_{\eta'}' +i \gamma_{\eta'}/2 ) (t-t'-t_0)}
    \notag\\
    && -
    [ C_{\vec{q} \eta} (t_0), C_{\vec{q}' \eta'}^\dagger (t-t') ]
        e^{-i( \omegatil_{\eta} -i \gamma_\eta/2 ) (t-t_0)}
    \notag\\
    && -
    [ C_{\vec{q} \eta} (t), C_{\vec{q}' \eta'}^\dagger (t_0) ]
        e^{i( \omegatil_{\eta'}' +i \gamma_{\eta'}/2 ) (t-t'-t_0)}
    \notag\\
    && +
    [ C_{\vec{q} \eta} (t), C_{\vec{q}' \eta'}^\dagger (t-t') ].
\end{eqnarray}
The commutation relations for $b_{\vec{q} \eta} (t_0)$ and $c_{\vec{q} \eta} (t_0)$ lead to
\begin{eqnarray}
    [ \bar{b}_{\vec{q} \eta} (t), \bar{b}_{\vec{q}' \eta'}^\dagger (t-t')]
    =
    \delta_{\vec{q}, \vec{q}'}
    \delta_{\eta, \eta'}
    &\Bigg[& \left(
        1 + \sum_r |\zeta_{r\eta}|^2 \frac{1}{(\omega_{r\eta} -\omegatil_\eta)^2 +\gamma_\eta^2/4}
    \right)
    e^{-i\omegatil_\eta t'} e^{-\gamma_\eta (2t+2t_0-t')/2}
    \notag\\
    &&
    - \sum_r |\zeta_{r\eta}|^2 \frac{
            e^{i \omega_{r\eta} (t-t'-t_0)}
        }{(\omega_{r\eta} -\omegatil_\eta)^2 +\gamma_\eta^2/4}
        e^{-i( \omegatil_{\eta} -i \gamma_\eta/2 ) (t-t_0)}
    \notag\\
    &&
    - \sum_r |\zeta_{r\eta}|^2 \frac{
            e^{-i \omega_{r\eta} (t-t_0)}
        }{(\omega_{r\eta} -\omegatil_\eta)^2 +\gamma_\eta^2/4}
        e^{i( \omegatil_\eta +i \gamma_\eta/2 ) (t-t'-t_0)}
    \notag \\
    &&
    + \sum_r |\zeta_{r\eta}|^2 \frac{
            e^{-i \omega_{r\eta} t'}
        }{(\omega_{r\eta} -\omegatil_\eta)^2 +\gamma_\eta^2/4}
    \Bigg] .
\end{eqnarray}

Note that we have sums of frequencies, but to be consistent with the assumptions in the derivation of the phonon dynamics solution $\bar{b}_{\vec{q}\eta}(t)$ which considers a continuum of frequencies in the reservoir \cite{kasperczyk2015, vento2025, guimaraes2020},
we must take
$\sum_r |\zeta_{r\eta}|^2 \rightarrow \int_0^\infty d{\omega_{r\eta}} \nu(\omega_{r\eta}) |\zeta_\eta(\omega_{r\eta})|^2 $.
Then, using $\nu(\omega_{r\eta}) |\zeta_\eta(\omega_{r\eta})|^2 = \gamma_\eta /(2\pi)$, and that $\omegatil_\eta/\gamma_\eta \gg 1$, which allows to take the lower limit of the integral to $-\infty$,
the commutator can be written as
\begin{eqnarray}
    [ \bar{b}_{\vec{q} \eta} (t), \bar{b}_{\vec{q}' \eta'}^\dagger (t-t')]
    =
    \delta_{\vec{q}, \vec{q}'}
    \delta_{\eta, \eta'}
    &\Bigg[& \left(
        1 + \int_{-\infty}^\infty \frac{\gamma_\eta}{2\pi}
        \frac{1}{(\omega_{r\eta} -\omegatil_\eta)^2 +\gamma_\eta^2/4}
        d\omega_{r\eta}
    \right)
    e^{-i\omegatil_\eta t'} e^{-\gamma_\eta (2t+2t_0-t')/2}
    \notag\\
    &&
    - \left( \int_{-\infty}^\infty \frac{\gamma_\eta}{2\pi} \frac{
            e^{i \omega_{r\eta} (t-t'-t_0)}
        }{(\omega_{r\eta} -\omegatil_\eta)^2 +\gamma_\eta^2/4}
        d\omega_{r\eta} \right)
        e^{-i( \omegatil_{\eta} -i \gamma_\eta/2 ) (t-t_0)}
    \notag\\
    &&
    - \left( \int_{-\infty}^\infty \frac{\gamma_\eta}{2\pi} \frac{
            e^{-i \omega_{r\eta} (t-t_0)}
        }{(\omega_{r\eta} -\omegatil_\eta)^2 +\gamma_\eta^2/4}
        d\omega_{r\eta} \right)
        e^{i (\omegatil+i\gamma_\eta/2) (t-t'-t_0)}
    \notag \\
    &&
    + \left( \int_{-\infty}^\infty \frac{\gamma_\eta}{2\pi} \frac{
            e^{-i \omega_{r\eta} t'}
        }{(\omega_{r\eta} -\omegatil_\eta)^2 +\gamma_\eta^2/4}
        d\omega_{r\eta} \right)
    \Bigg] ,
\end{eqnarray}
which then evaluates to
\begin{eqnarray}
    [ \bar{b}_{\vec{q} \eta} (t), \bar{b}_{\vec{q}' \eta'}^\dagger (t-t')]
    =
    \delta_{\vec{q}, \vec{q}'}
    \delta_{\eta, \eta'}
    &\bigg[&
    2 e^{-i\omegatil_\eta t'} e^{-\gamma_\eta (2t+2t_0-t')/2}
    - e^{i \omegatil (t-t'-t_0)} e^{- \gamma_\eta |t-t_0-t'|/2}
        e^{-i (\omegatil-i\gamma_\eta/2) (t-t_0)}
    \notag\\
    &&
    - e^{-i \omegatil (t-t_0)} e^{- \gamma_\eta |t-t_0|/2}
        e^{i (\omegatil+i\gamma_\eta/2) (t-t'-t_0)}
    + e^{-i \omegatil t'} e^{- \gamma_\eta |t'|/2}
    \bigg] .
\end{eqnarray}

The expression above simplifies to Eq. (\ref{eq:b_commutator}) of the main article,
\begin{equation}
	[ \bar{b}_{\vec{q} \eta} (t), \bar{b}_{\vec{q}' \eta'}^\dagger (t-t')]
    =
    e^{-i \omegatil_\eta t'}
    e^{-\gamma_\eta |t'| /2}
    \delta_{\vec{q},\vec{q}'}
    \delta_{\eta,\eta'}.
\end{equation}
Interestingly, $[ \bar{b}_{\vec{q} \eta} (t), \bar{b}_{\vec{q}' \eta'}^\dagger (t-t')] = [ C_{\vec{q} \eta} (t), C_{\vec{q}' \eta'}^\dagger (t-t')]$, and the same result is obtained if one works only with the stationary part of the solution, with the transient term cancelling out.

\section{From spatial to temporal frequency susceptibility}\label{app:susceptibility}

Because our theory is fully spatio-temporal, we have to go through a spatial third-order susceptibility in order to derive the usual temporal frequency susceptibility due to vibrational Raman scattering.
We define by $\chi^{(3) \T{R}}_{ijj'i'} (-\vec{k}_A, \vec{k}_\ell, \vec{k}_\ell', -\vec{k}_S)$
the third-order spatial susceptibility due to the vibrational Raman processes scattering photons in the anti-Stokes mode $\vec{k}_A$ with frequency $\omega_A$,
associated with the absorption of two laser photons in modes $\vec{k}_\ell$ and $\vec{k}_\ell'$, frequencies $\omega_\ell$ and $\omega_\ell'$ respectively, and the creation of a Stokes photon in mode $\vec{k}_S$ with frequency $\omega_S$.
It can be identified in the expression
\begin{equation}
    P^{\T{R}(1)}_i(\vec{k}_A,\omega_A)
    =
    \sum_{\vec{k}_\ell} \sum_{\vec{k}_\ell'} \sum_{\vec{k}_S}
    \chi^{(3) \T{R}}_{ijj'i'} (-\vec{k}_A, \vec{k}_\ell, \vec{k}_\ell', -\vec{k}_S)
    \mc{E}_{\vec{k}_\ell j} \mc{E}_{\vec{k}_\ell' j'}
	E_{\vec{k}_S i'}^\dagger
    .
\end{equation}
In order to obtain it, we take the spatio-temporal Fourier transform of the $AS$ term of Eq. (\ref{eq:pol_fAS}) of the main article, which has the electric polarization in the anti-Stokes mode, and the Stokes photon in the correlated field operator,
\begin{equation}
    P^{\T{R}(1)}_i(\vec{k}_A,\omega_A)
    =
    \int_\mc{V}
    \int_{-\infty}^{\infty}
        \left[ P^{\T{R}(1)}_i(\vec{r},t) \right]^{AS}_{++-}
    e^{-i(\vec{k}_A\cdot\vec{r} - \omega_A t)}  dt d^3\vec{r}
    .
\end{equation}
With the use of $\int_\mc{V} e^{i(\vec{k}-\vec{k}')\cdot\vec{r}} d^3\vec{r} = V_S \delta_{\vec{k},\vec{k}'}$
and
$f_{\pm \pm \pm}^{(\frac{S}{A}) (\frac{S}{A})} (\omega)
\equiv \int_{-\infty}^{\infty}
f_{\pm \pm \pm}^{(\frac{S}{A}) (\frac{S}{A})} (t)
e^{i\omega t} dt$,
we can write
\begin{equation}
    P^{\T{R}(1)}_i(\vec{k}_A,\omega_A)
    =
	\sum\limits_{\vec{k}_\ell} \sum\limits_{\vec{k}_\ell'}
    \sum\limits_{\vec{q}}
    \sum_\eta \frac{\gamma_\eta}{2\pi}
    \bar{\alpha}_{ij,\eta}(\vec{q}) \bar{\alpha}_{i'j',\eta}^*(\vec{q})
    f_{++-}^{AS}(\omega_A)
    \delta_{\vec{k}_A, \vec{k}_\ell+\vec{q}}
    \mc{E}_{\vec{k}_\ell j} \mc{E}_{\vec{k}_\ell' j'}
	E_{(\vec{k}_\ell'-\vec{q}) i'}^\dagger
    ,
\end{equation}
where
\begin{equation}
    \bar{\alpha}_{ij,\eta}(\vec{q}) \equiv
    \left( \frac{2\pi}{\gamma_\eta}
        \frac{V_S}{2 M_{\eta} \omega_{\eta 0}}
    \right)^{1/2}
    N\epsilon_0
    \alpha_{ijm}
    \varepsilon_{\vec{q} \eta m}
\end{equation}
is the tensorial quantity that carries the polarization properties of the scattering.

The sum of the scattered field is on the phonon wave vector $\vec{q}$,
and by writing $\vec{k}_S \equiv \vec{k}_\ell'-\vec{q}$, it can be recast as
\begin{eqnarray}
    P^{\T{R}(1)}_i(\vec{k}_A,\omega_A)
    =
	\sum\limits_{\vec{k}_\ell} \sum\limits_{\vec{k}_\ell'}
    \sum\limits_{\vec{k}_S}
    \sum_\eta&& \frac{\gamma_\eta}{2\pi}
    \bar{\alpha}_{ij,\eta}(\vec{k}_\ell'-\vec{k}_S) \bar{\alpha}_{i'j',\eta}^*(\vec{k}_\ell'-\vec{k}_S)
    f_{++-}^{AS}(\omega_A)
    \notag\\
    &&\times
    \delta_{\vec{k}_\ell+\vec{k}_\ell', \vec{k}_S+\vec{k}_A}
    \mc{E}_{\vec{k}_\ell j} \mc{E}_{\vec{k}_\ell' j'}
	E_{\vec{k}_S i'}^\dagger
    ,
\end{eqnarray}
such that, by inspection,
\begin{equation}
    \chi^{(3) \T{R}}_{ijj'i'} (-\vec{k}_A, \vec{k}_\ell, \vec{k}_\ell', -\vec{k}_S)
    =
    \sum_\eta \frac{\gamma_\eta}{2\pi}
    \bar{\alpha}_{ij,\eta}(\vec{k}_\ell'-\vec{k}_S) \bar{\alpha}_{i'j',\eta}^*(\vec{k}_\ell'-\vec{k}_S)
    f_{++-}^{AS}(\omega_A)
    \delta_{\vec{k}_\ell+\vec{k}_\ell', \vec{k}_S+\vec{k}_A}
    .
\end{equation}
The Kronecker delta $\delta_{\vec{k}_\ell+\vec{k}_\ell', \vec{k}_S+\vec{k}_A}$ is the momentum conservation condition in the FWM process.

A closer look at $f_{++-}^{AS} (\omega_A)$, with $\omega_S \equiv \omega(\vec{k}_S)$,
\begin{eqnarray}\label{eq:fAS_freq}
    f_{++-}^{AS} (\omega_A; \omega_\ell, \omega_\ell', \omega_S)
    &=&
    \frac{
        2\pi \delta( \omega_{\ell} + \omega_{\ell}' - \omega_{S} -\omega_A)
    }
    {( \omega_{\ell}' - \omega_{S} - \omegatil_\eta + i \gamma_\eta/2
    )}
    \notag\\
    &&-i \frac{ 1 }
    {( \omega_{\ell}' - \omega_{S} - \omegatil_\eta + i \gamma_\eta/2)
    ( \omega_{\ell} -\omega_A + \omegatil_\eta -i \gamma_\eta/2)
    },
\end{eqnarray}
reveals two terms.
The first one carries the energy conservation condition in the Dirac delta, while the second one is composed of two scattering probability amplitudes, one from $\vec{k}_\ell'$ to $\vec{k}_S$ and another from $\vec{k}_\ell$ to $\vec{k}_A$.
In the second term, however, no conservation condition is imposed in the overall event, with the two scattering processes that compose the FWM being independent from each other.
Since the energy conserving term has an infinite peak at $\omega_A = \omega_\ell + \omega_\ell' -\omega_S$, any other probability amplitude in the second term, in which $\omega_A \neq \omega_\ell + \omega_\ell' -\omega_S$, will be negligible in comparison with the first one.
This is a feature of the fact that we are in the frequency domain, and the long integration time needed to obtain it cause the second term in Eq. (\ref{eq:f_time}) of the main article, which contains an $e^{-\gamma_\eta t/2}$ factor, to vanish in comparison with the first one, which does not contain it.
Therefore, both momentum and energy conservation conditions are present in this more general spatial susceptibility.

We can further simplify the expression if we restrict ourselves to forward scattering and degenerate phonon modes (i.e. $M_\eta = M$, $\omega_{\eta 0} = \omega_\eta$, and $\gamma_\eta = \gamma$).
In this case $\bar{\alpha}_{ij,\eta}$ is independent of the phonon wave vector, and the momentum and energy conditions are completely separated.
We can then, finally, define the third-order susceptibility $\chi^{(3) \T{R}}_{ijj'i'} (-\omega_A, \omega_\ell, \omega_\ell', -\omega_S)$ by
\begin{equation}
    \chi^{(3) \T{R}}_{ijj'i'} (-\vec{k}_A, \vec{k}_\ell, \vec{k}_\ell', -\vec{k}_S)
    \equiv \chi^{(3) \T{R}}_{ijj'i'} (-\omega_A, \omega_\ell, \omega_\ell', -\omega_S)
    \delta_{\vec{k}_\ell+\vec{k}_\ell', \vec{k}_S+\vec{k}_A},
\end{equation}
such that,
with $
A^{\T{R}}_{ijj'i'} \equiv
    \sum_\eta
    \bar{\alpha}_{ij,\eta} \bar{\alpha}_{i'j',\eta}^*
    $,
\begin{equation}
    \chi^{(3) \T{R}}_{ijj'i'} (-\omega_A, \omega_\ell, \omega_\ell', -\omega_S)
    =
    A^{\T{R}}_{ijj'i'}
    \frac{ \gamma }
    {( \omega_{\ell}' - \omega_S - \omegatil + i \gamma/2
    )}
    \delta( \omega_{\ell} + \omega_{\ell}' - \omega_{S} -\omega_A)
\end{equation}
is the susceptibility due to the FWM Raman interaction in the material,
for the $\omega_\ell'$ mode scattering into $\omega_S$, and $\omega_\ell$ into $\omega_A$.
It has the same functional form as the susceptibility calculated for stimulated Raman scattering \cite{bloembergen, boyd}, but since we have made a fully quantum treatment,
we have shown that this expression can be used even in the spontaneous SaS photon pair generation, as it has been without a formal derivation \cite{vento2025, sier2024}.

\end{document}